\documentclass[twocolumn,amsmath,amssymb,english,superscriptaddress,nofootinbib]{revtex4-1}

\usepackage{dcolumn}
\usepackage{babel}

\makeatother

\usepackage{graphicx}
\usepackage{amsmath}
\usepackage{bm}
\usepackage{multirow}

\begin{document}

\title{
Structure of even-even Cadmium isotopes from the beyond-mean-field
interacting boson model
}

\date{\today}

\author{K. Nomura}

\affiliation{Physics Department, Faculty of Science, University of Zagreb, HR-10000 Zagreb, Croatia}

\author{J. Jolie}

\affiliation{Institut f\"{u}r Kernphysik, Universit\"{a}t zu K\"{o}ln, D-90937 K\"{o}ln, Germany}

\begin{abstract}
The structure of even-even $^{108-116}$Cd isotopes is investigated
based on the self-consistent mean-field approach. 
By mapping the
 quadrupole-$(\beta,\gamma)$  
 deformation energy surface, obtained from the constrained self-consistent
 mean-field calculations with a choice of the Skyrme force and pairing
 property, onto the Hamiltonian of the interacting boson model with 
 configuration mixing, the strength parameters of the Hamiltonian are
 determined. 
The low-lying excitation spectra and electric quadrupole and monopole transition rates
 for the considered Cd nuclei are computed by the resultant Hamiltonian,
 and are compared in detail with the experimental data. 
Our semi-microscopic prediction identifies several intruder states as
 suggested empirically, and overall, provides a reasonable qualitative description
 of the experimental energy  levels and transition rates. 
\end{abstract}

\maketitle
\section{Introduction}

The $Z=50$ mass region
is very favourable for nuclear structure studies
due to the large abundance of stable isotopes
combined with the interesting features
of the nearby $Z=50$ proton shell closure
and with the occurrence of neutrons
in the middle of the $N=50$--$82$ shell.

The earliest work started with the observation
by Schraff-Goldhaber and Wesener~\citep{001}
that the Cd isotopes exhibit low-lying states
that resemble the quadrupole-vibrational excitations
of a surface with spherical equilibrium
as predicted by the collective model of Bohr and Mottelson~\citep{002}.
Besides the one-phonon quadrupole state,
candidates for two-phonon quadrupole states
were observed around 1.2~MeV,
twice the energy of the first-excited $2^+$ state.
The two-phonon states were found non-degenerate,
indicating the need to include anharmonic effects
in the phonon--phonon interactions.
Furthermore, additional $0^+$ and $2^+$ states
were observed in transfer studies~\citep{cohen1961}.
Attempts by Bes and Dussel to explain these
as strongly anharmonic three-phonon states failed~\citep{BES69}.
The explanation of the additional states was then related
to two-particle--two-hole (2p--2h) excitations
of the protons across the $Z=50$ closed shell.
Evidence that the extra $0^+$ and $2^+$ states 
were indeed 2p--4h states
was obtained from ($^3$He,n)
two-proton transfer experiments~\citep{AND77}.
By the early nineties intruder bands were identified
in most even--even Cd isotopes. 
Strong support for the intruder interpretation
came from the systematic behaviour of these states
as a function of the number of valence neutrons.
Due to the increase in neutron--proton quadrupole interaction,
intruder states decrease in energy
proportional to the number of neutrons,
reaching a lowest value near mid-shell~\citep{HEY87}.
In the stable Cd isotopes this mechanism
gives rise to intruder and two-phonon states close in energy,
resulting in complex spectra.

The even-even Cd isotopes have been studied intensively from $N=46$
~\citep{DAV17} till $N=86$ ~\citep{LOR15}. The most detailed information
is obtained for the stable isotopes with $A=106$ till $A=116$
\citep{FAH88,135Jolie01, 
136Jolie02,137Jolie03,138Jolie04,139Jolie05,140Jolie06,141Jolie07,142Jolie08,143Jolie09,144Jolie10,145Jolie11,
146Jolie12,147Jolie13,148Jolie14,152Jolie22_not_listed,013_Boelaert2,133free,134free,015,016,017,018,019,020,021,022,023,025,026,027,
028,029,030,031,032,033,034,035,036,037,038,039,040,041,042,043,044,045,046,047}). 
The systematics for those data are discussed in Refs. \citep{048,049,050_NNDC}. A recent review on the structure
of $^{100}$Sn and neighboring nuclei including the light Cd isotopes below $N=50$ is given in \citep{051}.

Shell-model calculations have been performed for the lighter Cd isotopes \citep{006,007,008,009,010,011,012_Boelaert1,014,016,052} up to $^{108}$Cd \citep{SCH17}. The used model spaces, however, do not consider proton $2p-2h$ excitation across $Z=50$.
Different calculation that include the intruder states have been
performed using the Interacting Boson Model in its simplest version with
$s$ and $d$ bosons (IBM-1) or more elaborated versions with $s$, $p$, $d$ and
$f$ bosons, broken pairs or using the neutron-proton version (IBM-2)\citep{015,016,019,020, 021,097,136Jolie02,149Jolie16,098,099,101,100,150Jolie21,112,152Jolie22_not_listed,024}.
Besides that, other studies, starting from a general collective Bohr
Hamiltonian, derived from a microscopic starting point using a Skyrme
force, calculations using the Adiabatic Time-Dependent Hartree-Fock-Bogoliubov
(ATDHFB) method (for the nuclei $^{106-116}$Cd) \citep{054}, as
well as using a self-consistent HFB approach, starting from the finite
range Gogny interaction \citep{055} have been carried out.

In this work we want to use the results of the self-consistent mean-field approach
to perform IBM-2 calculations with normal and intruder states for the $A=108-116$ even-even Cd
isotopes using the approach introduced in \citep{nomura2012sc}. A
detailed comparison with the extensive experimental data set on energy
levels and electric quadrupole and monopole transitions is then made.

\section{Theoretical framework}

In this section we briefly outline the theoretical scheme used
in the present work to study the even-even $^{108-116}$Cd isotopes. For
the detailed accounts of the method, the reader is referred to 
Refs.~\citep{nomura2008,nomura2012sc}. 

We have first carried out, for each Cd nucleus, the
constrained self-consistent mean-field (SCMF) calculations  to obtain the
deformation energy surface in the $(\beta,\gamma)$ quadrupole deformation
space. The constraints imposed are on mass quadrupole
moments $Q_{20}$ and $Q_{22}$, which can be associated with the axial  $\beta$
and triaxial $\gamma$ deformation parameters of the collective model \citep{002}. 
For the SCMF calculations, we have employed the Hartree-Fock+BCS method \citep{ev8,ev8r}, where
the particle-hole interaction is modeled by the SLy6 parametrization \citep{sly6} of 
the Skyrme force and the particle-particle channel is described by the
density-dependent zero-range pairing force with the strength of $V_0=$1000 MeV
fm$^3$  truncated below and above the Fermi surface by 5 MeV, for both protons and neutrons.
More details about the HF+BCS calculation can be found in Refs.~\citep{ev8,ev8r}.

\begin{figure}[htb!]
\begin{center}
\includegraphics[width=\columnwidth]{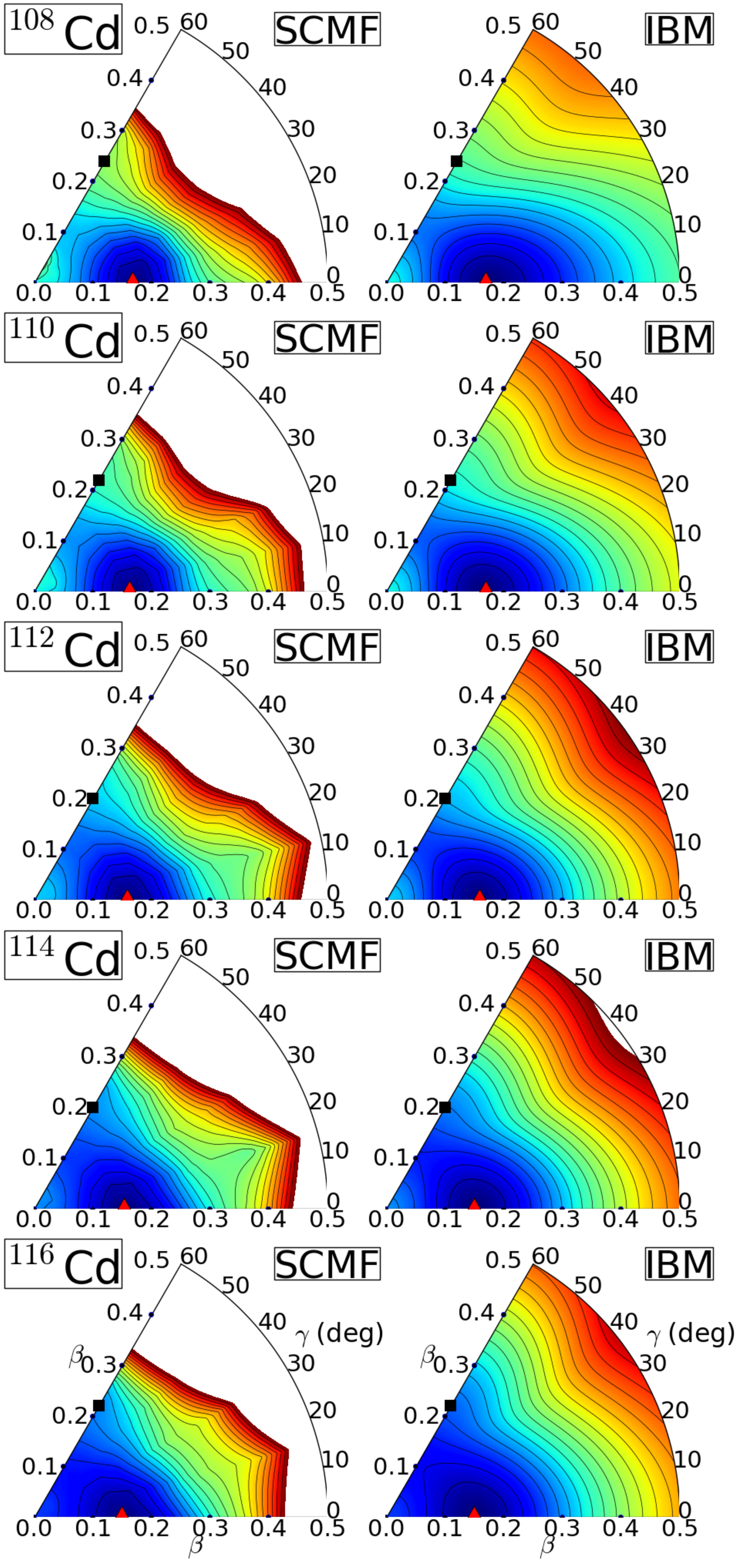}
\caption{(Color online) Left column: Contour plots for the $\beta\gamma$ deformation energy
 surfaces for $^{108-116}$Cd, that have been obtained from the self-consistent
 mean-field (SCMF) calculation using
 the Skyrme SLy6 parametrization with the density-dependent zero-range pairing
 interaction with the strength $V_0=1000$ MeV fm$^{3}$. Right column: the corresponding IBM-2 energy
 surfaces. Energy difference between neighbouring contours is 250
 keV. The global minimum is indicated by solid triangle, while the local
 minimum is identified by solid squares.} 
\label{fig:pes}
\end{center}
\end{figure}

\begin{figure}[htb!]
\begin{center}
\includegraphics[width=\columnwidth]{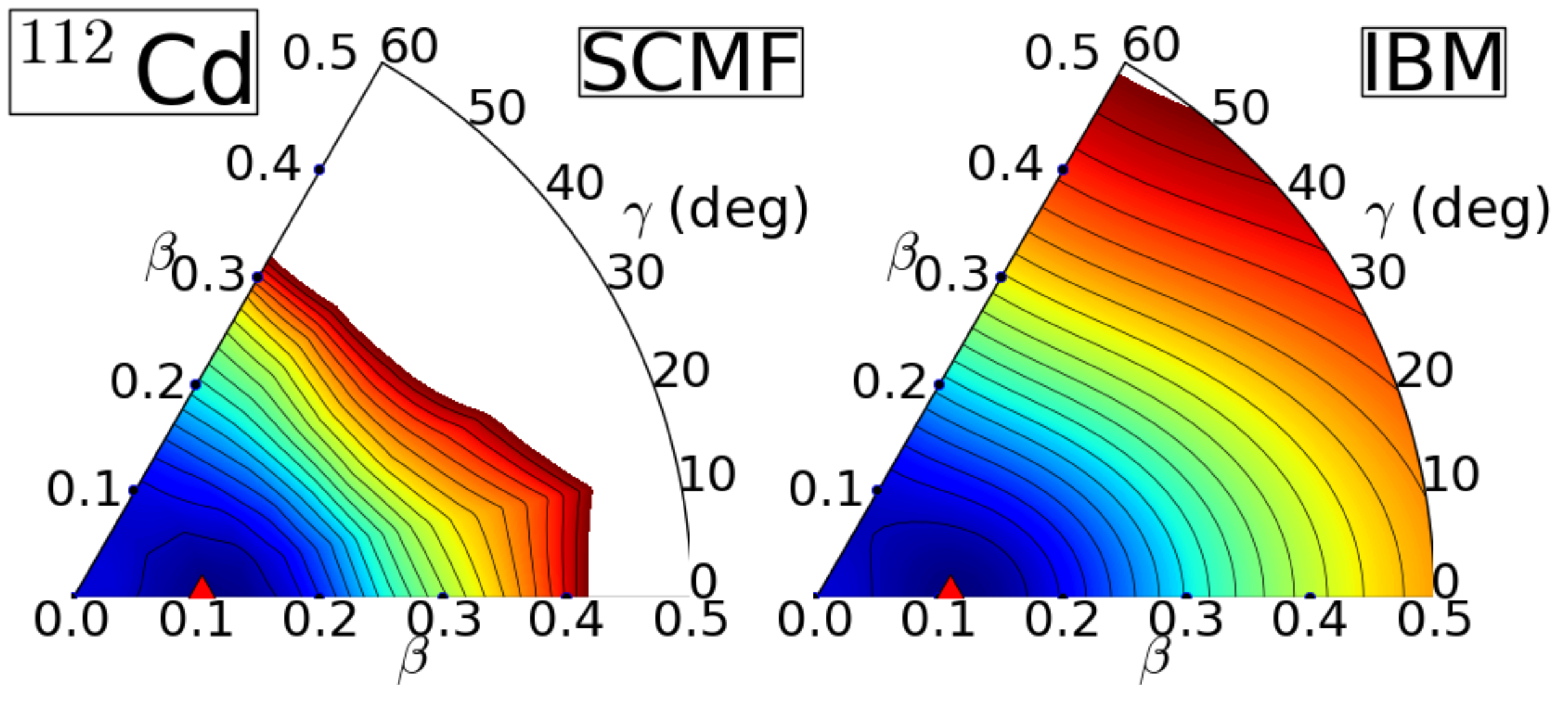}
\caption{(Color online) Same as Fig.~\ref{fig:pes}, but for the
 calculation with the pairing strength of $V_0=1250$ MeV fm$^3$ for the $^{112}$Cd.} 
\label{fig:pes-pairing}
\end{center}
\end{figure}

On the left hand side of Fig.~\ref{fig:pes} we draw contour plots of the
$(\beta,\gamma)$-deformation energy surfaces for the $^{108-116}$Cd
isotopes, obtained from the above-mentioned SCMF calculation. 
For all the considered Cd nuclei, a prolate global minimum is found with
moderate axial deformation $\beta\approx 0.15$. 
We also observe on the oblate side ($\gamma\approx 60^{\circ}$) a much
less pronounced local minimum between $\beta=0.2$ and 0.3. 
In order to examine the sensitivity of the SCMF result to the pairing
property, we depict, on the left-hand side of
Fig.~\ref{fig:pes-pairing}, the SCMF energy surface for the $^{112}$Cd
nucleus, obtained with the same Skyrme force but with the pairing
strength of $V_0=1250$ MeV fm$^3$. 
By comparing it with the corresponding SCMF energy surface in
Fig~\ref{fig:pes} in the case of $V_0=1000$ MeV fm$^3$, one could notice that,
when the pairing strength is increased, the surface has a less pronounced
prolate minimum and more resembles the potential typical of spherical
vibrator. In addition, local minimum is
no longer visible on the oblate side of the SCMF energy surface with
$V_0=1250$ MeV fm$^3$. 
Later we show how the difference in the SCMF energy surface between the
different pairing strengths influences the energy spectra.

The next step is to construct from those SCMF results configuration
mixing IBM-2 Hamiltonian by using the procedure of Ref.~\citep{nomura2012sc}. 
It is based on the method developed in \citep{nomura2008}, in which the SCMF
energy surface is mapped onto the expectation value of 
the IBM-2 Hamiltonian in the boson coherent state \citep{ginocchio1980}
so as to determine the strength parameters of the Hamiltonian. 
In contrast to many phenomenological IBM studies, there is no adjustment
of the parameters to experimental data in this procedure. 
Diagonalization of the resulting Hamiltonian in the laboratory frame
provides one with excitation spectra and electromagnetic transition
rates.

The IBM-2 is comprised of the neutron (proton) $s_{\nu}$ ($s_{\pi}$) and
$d_\nu$ ($d_\pi$) bosons, which 
represent, respectively, the collective $J=0^+$ and $2^+$ pairs of
valence neutrons (protons) \citep{otsuka1978}. 
The number of neutron (proton) bosons, denoted as $N_\nu$
($N_{\pi}$), is equal to that of the valence neutron 
particles/holes (proton holes). 
Here we take the doubly-magic nuclei $^{100}$Sn and $^{132}$Sn (for $^{116}$Cd)  as inert cores for the
bosons. In addition, by using the procedure proposed by Duval and Barrett to
incorporate intruder states in the IBM \citep{duval1981}, we
take into account the proton $2p-2h$ intruder excitation 
across the $Z=50$ shell closure. 
In this proposal \citep{duval1981} particle-like and hole-like bosons are
not distinguished and, as the excitation of a pair (or
boson) increases the boson number by two, the $0p-0h$ and $2p-2h$
configurations differ in boson number by two. 
Applying this boson-number counting rule to the considered $^{106-116}$Cd nuclei,
$N_{\pi}=1$ and 3 for the normal ($0p-0h$) and intruder ($2p-2h$) 
configurations, respectively, while $5\leq N_{\nu}\leq 8$. The
configuration mixing IBM-2 Hamiltonian is then written as: 
\begin{eqnarray}
\label{eq:ham-cm}
 \hat H = {\cal\hat P}_1\hat H_1{\cal\hat P}_1 + {\cal\hat P}_3(\hat H_3 + \Delta){\cal\hat P}_3 + \hat
  H_{\rm mix}, 
\end{eqnarray}
where $\hat H_1$ ($\hat H_3$) and ${\cal\hat P}_1$ (${\cal\hat P}_3$) are the
Hamiltonian of and the projection operator onto the normal and intruder
configuration spaces, respectively. $\Delta$ stands for the energy
needed to promote a proton boson across the shell closure. 
$\hat H_{\rm mix}$ in the above equation is the term that is allowed to
mix the two configurations.
Here we employ the same Hamiltonian as in \citep{NOM13}, but the three-body
boson term is not included here.

The coherent state for the configuration mixing IBM was introduced in
\citep{frank2004} as the direct sum of the coherent state for each
unperturbed configuration. 
The energy surface of the configuration-mixing IBM-2 is obtained as the
lower eigenvalue of the $2\times 2$ coherent-state matrix \citep{frank2004}. 
The analytical expressions for each component of the coherent-state
matrix are found in \citep{NOM13}. 

The parameters for the Hamiltonian for each configuration are determined
by associating the Hamiltonian with each mean-field minimum, i.e., $0p-0h$
Hamiltonian for the prolate minimum, and $2p-2h$ one for the oblate
local minimum, and the energy offset $\Delta$ and the mixing strength in $\hat H_{\rm
mix}$ are determined so that the energy difference between the two
mean-field minima and the barrier height for these minima, respectively,
are reproduced. 
The derived strength parameters for $^{108-116}$Cd are
listed in Table~\ref{tab:para}. 

\begin{table}
 \begin{center}
\caption{\label{tab:para}The parameters of the configuration mixing IBM-2 Hamiltonian in
   Eq.~(\ref{eq:ham-cm}) employed in the calculation (in MeV
   units). Their definitions are given in Ref.~\citep{NOM13}. For the mixing
   strength $\omega$ in $\hat H_{\rm mix}$ constant value of 
   $\omega=0.15$ MeV is used. }
  \begin{tabular}{llccccccc}
\hline\hline
 & & $\epsilon$ & $\kappa$ & $\chi_\nu$ & $\chi_\pi$ & $\kappa'$ & $\Delta$ \\
\hline
\multirow{2}{*}{$^{108}$Cd} & normal   & 0.297 & -0.490 & -0.446 & -0.950 & 0.0383 & \multirow{2}{*}{2.572} \\
           & intruder & 0.486 & -0.200 & 0.600 & 0.650 & 0.0157 &  &  \\
\multirow{2}{*}{$^{110}$Cd} & normal   & 0.404 & -0.513 & -0.269 & -0.845 & 0.0415 & \multirow{2}{*}{2.221} \\
           & intruder & 0.368 & -0.195 & 0.400 & 0.650 & 0.0420 &  &  \\
\multirow{2}{*}{$^{112}$Cd} & normal   & 0.497 & -0.495 & -0.127 & -0.845 & 0.0424 & \multirow{2}{*}{1.804} \\
           & intruder & 0.326 & -0.180 & 0.050 & 0.650 & 0.0490 &  &  \\
\multirow{2}{*}{$^{114}$Cd} & normal   & 0.462 & -0.417 & -0.126 & -0.844 & 0.0449 & \multirow{2}{*}{2.132} \\
           & intruder & 0.321 & -0.180 & 0.050 & 0.650 & 0.0498 &  &  \\
\multirow{2}{*}{$^{116}$Cd} & normal   & 0.626 & -0.422 & -0.217 & -0.696 & 0.0340 & \multirow{2}{*}{2.811} \\
           & intruder & 0.489 & -0.197 & 0.600 & 0.650 & 0.0302 &  &  \\
\hline
\end{tabular}
 \end{center}
\end{table}

On the right column of Figs.~\ref{fig:pes} and \ref{fig:pes-pairing} we
show the mapped IBM-2 energy surfaces. 
Note the intruder configuration has not been included for the
calculation of $^{112}$Cd with the pairing strength of 
$V_0=1250$ MeV fm$^3$, because there is no additional minimum on the
oblate side (see, Fig.~\ref{fig:pes-pairing}). 
One sees that the topology of the corresponding SCMF energy surfaces
in the neighborhood of the minimum is well reproduced by the IBM ones. 
The IBM surface, however, tends to be flat in the region far from the
minimum, compared to the SCMF one. 
Main reason is that we have paid particular attention to reproduce,
as much as possible, the topology of the SCMF energy surface in the
vicinity of, typically a few MeVs above, the minimum: The most relevant
mean-field configurations to low-lying quadrupole collective states are
those in the neighborhood of the minimum, while those
very far from it tend to be dominated by 
non-collective, i.e., quasiparticle, degrees of freedom, which are out
of the model space of the IBM-2 framework. 
Another reason is, of course, that the employed IBM-2 Hamiltonian and/or
coherent-state formalism may have been too
simple to account for every detail of the SCMF energy surface.

Using the resulting wave functions of the IBM-2 Hamiltonian, we have also computed the electric
quadrupole (E2) and monopole (E0) transition rates. We used the E2 operator: 
\begin{eqnarray}
\label{eq:e2}
 \hat T^{E2} = \sum_{\tau,i}{\cal\hat P}_i e^\tau_i\hat Q_{\tau,i}{\cal\hat P}_i,
\end{eqnarray}
where $\tau=\nu$ (neutron) or $\pi$ (proton), $i=1$ ($0p-0h$) or 3
($2p-2h$), $\hat Q_{\tau,i}$ is the quadrupole operator same as the one
in each unperturbed Hamiltonian, and $e^\tau_i$ the boson effective charge. 
The E0 operator used in this study is given by: 
\begin{eqnarray}
 \label{eq:e0}
 \hat T^{E0} = \sum_{\tau,i}{\cal \hat
  P}_i(\beta_{\tau,i}\hat n_{d\tau,i} + \gamma_{\tau,i}\hat
  N_{\tau,i}){\cal \hat P}_i, 
\end{eqnarray}
where $\hat n_{d\tau,i}$ is the  neutron or
proton $d$-boson number operator for a given configuration while $\hat
N_{\tau,i}$ the total neutron or proton boson 
number. 
$\beta_{\tau,i}$ and $\gamma_{\tau,i}$ are parameters.

\section{Comparison with experiment}

\begin{figure*}[htb!]
\begin{center}
\begin{tabular}{ccc}
\includegraphics[width=0.3\linewidth]{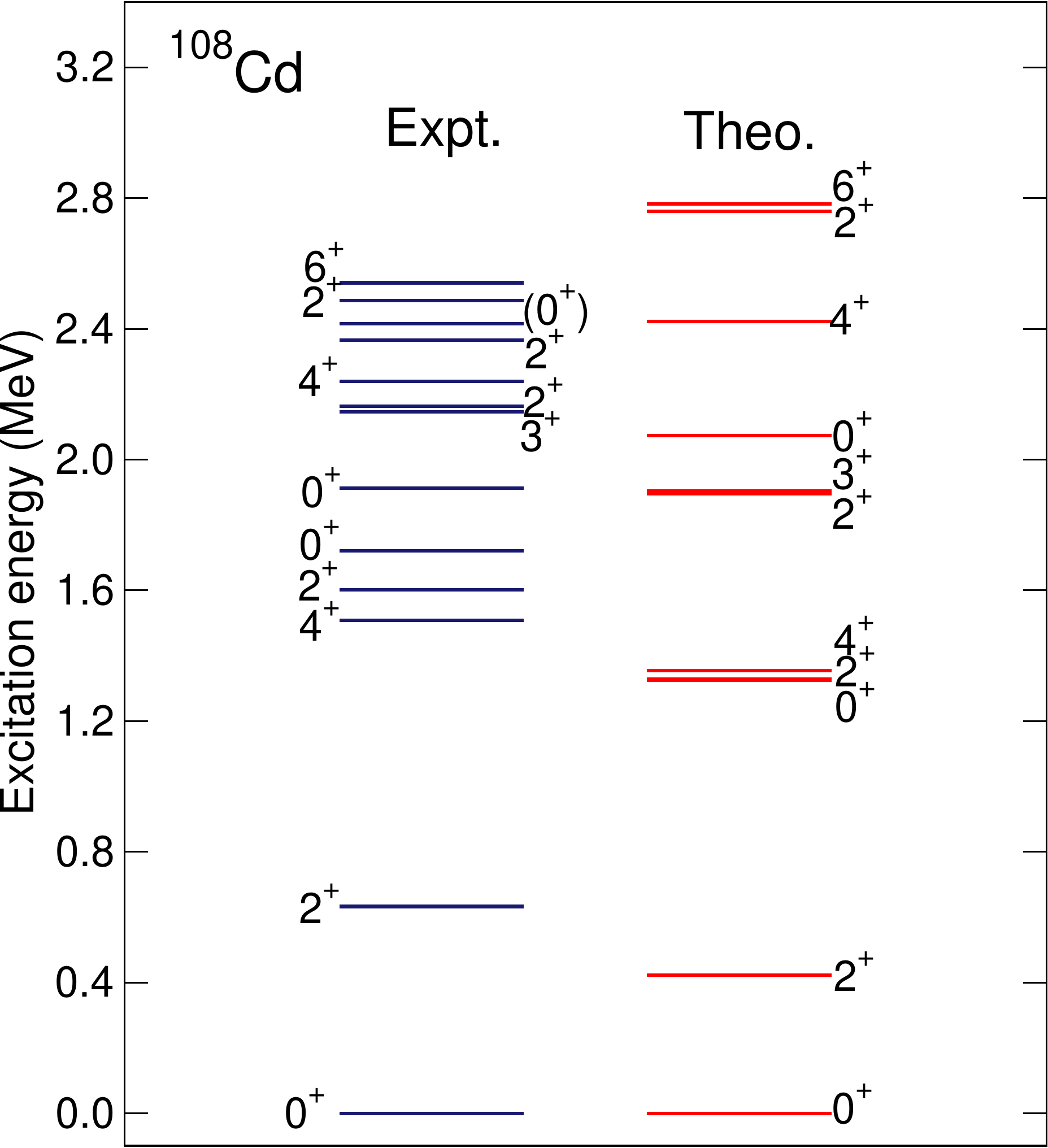} &
\includegraphics[width=0.3\linewidth]{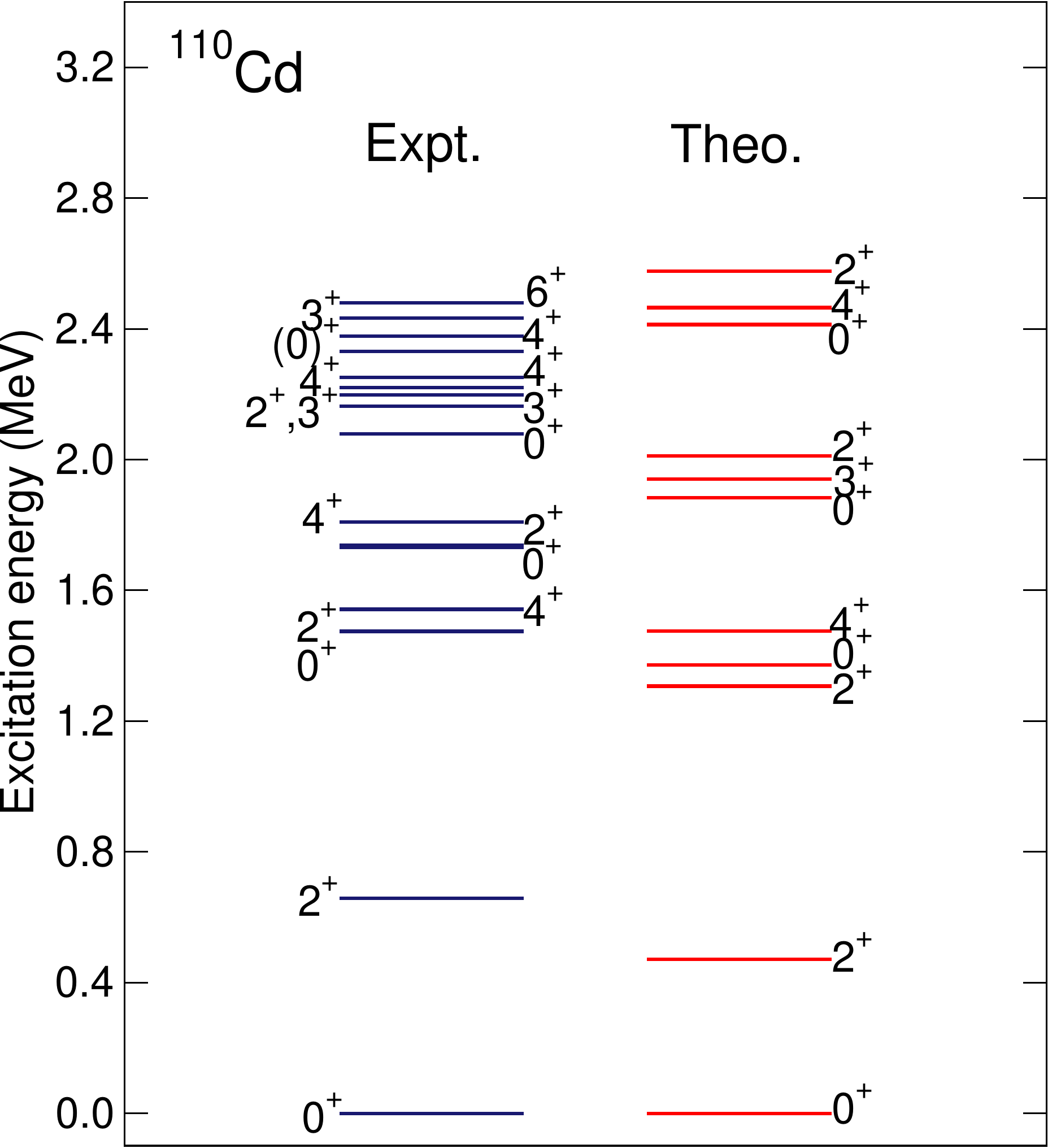} &
\includegraphics[width=0.3\linewidth]{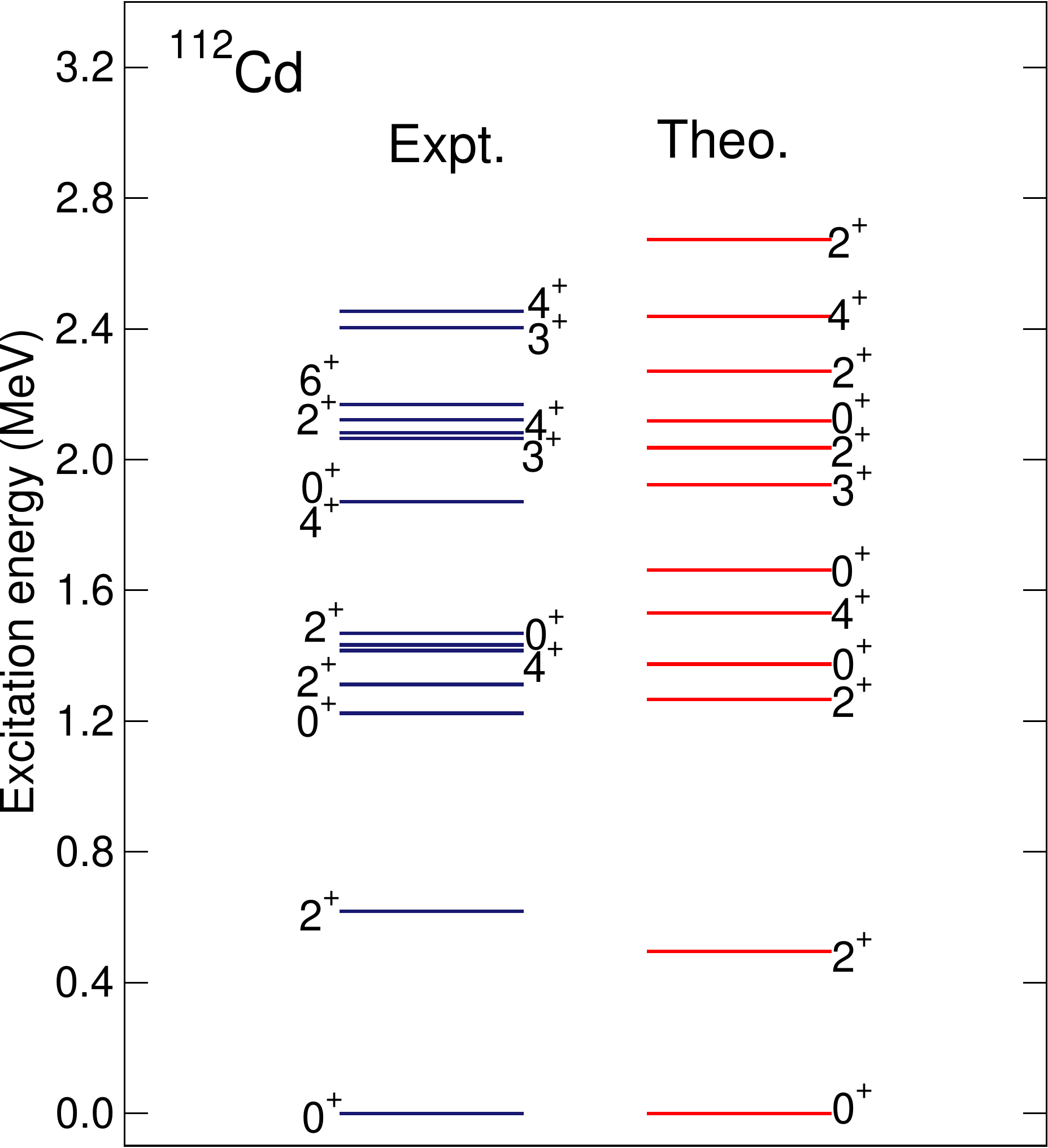} \\
\end{tabular}
\begin{tabular}{cc}
\includegraphics[width=0.3\linewidth]{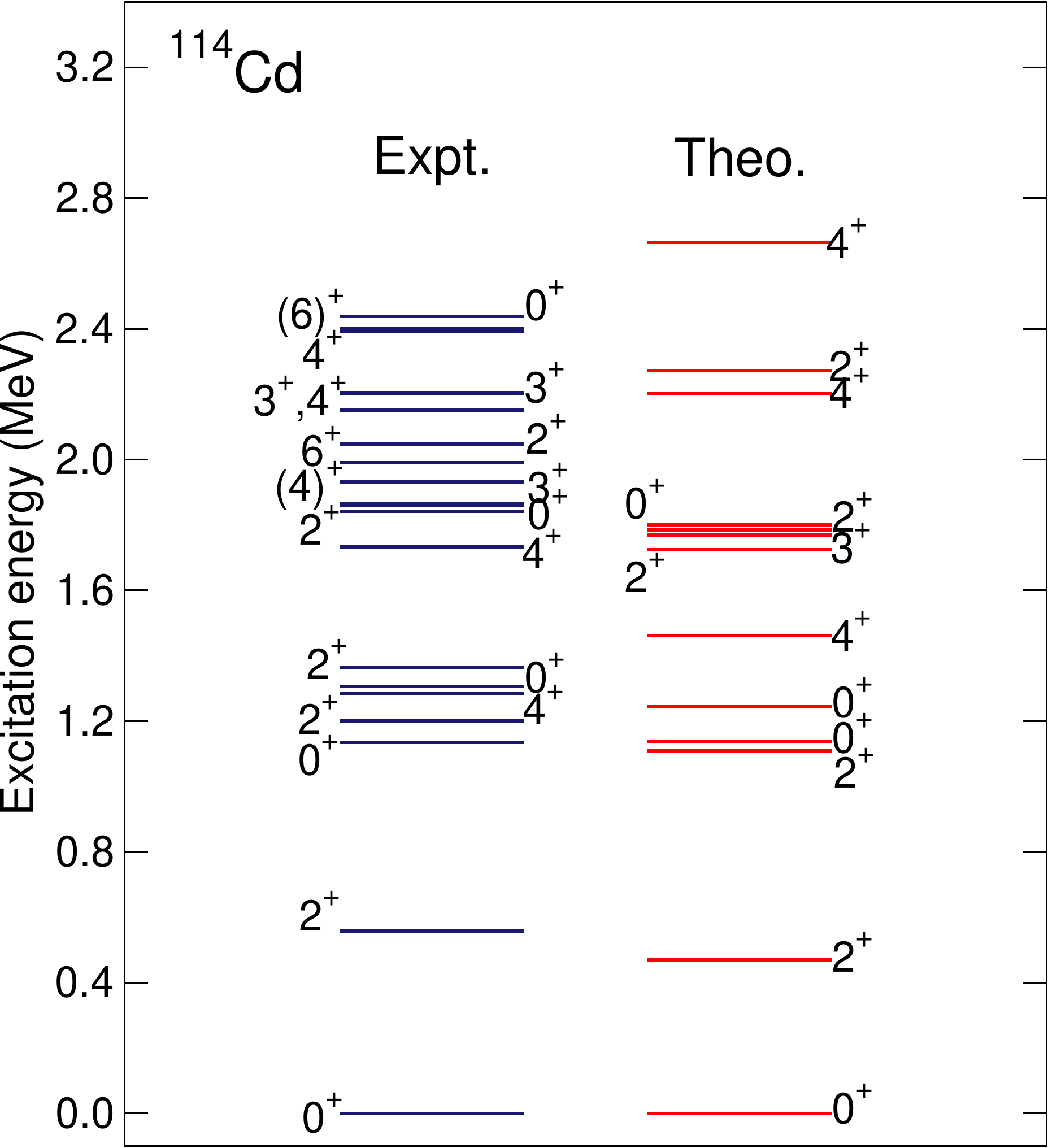} &
\includegraphics[width=0.3\linewidth]{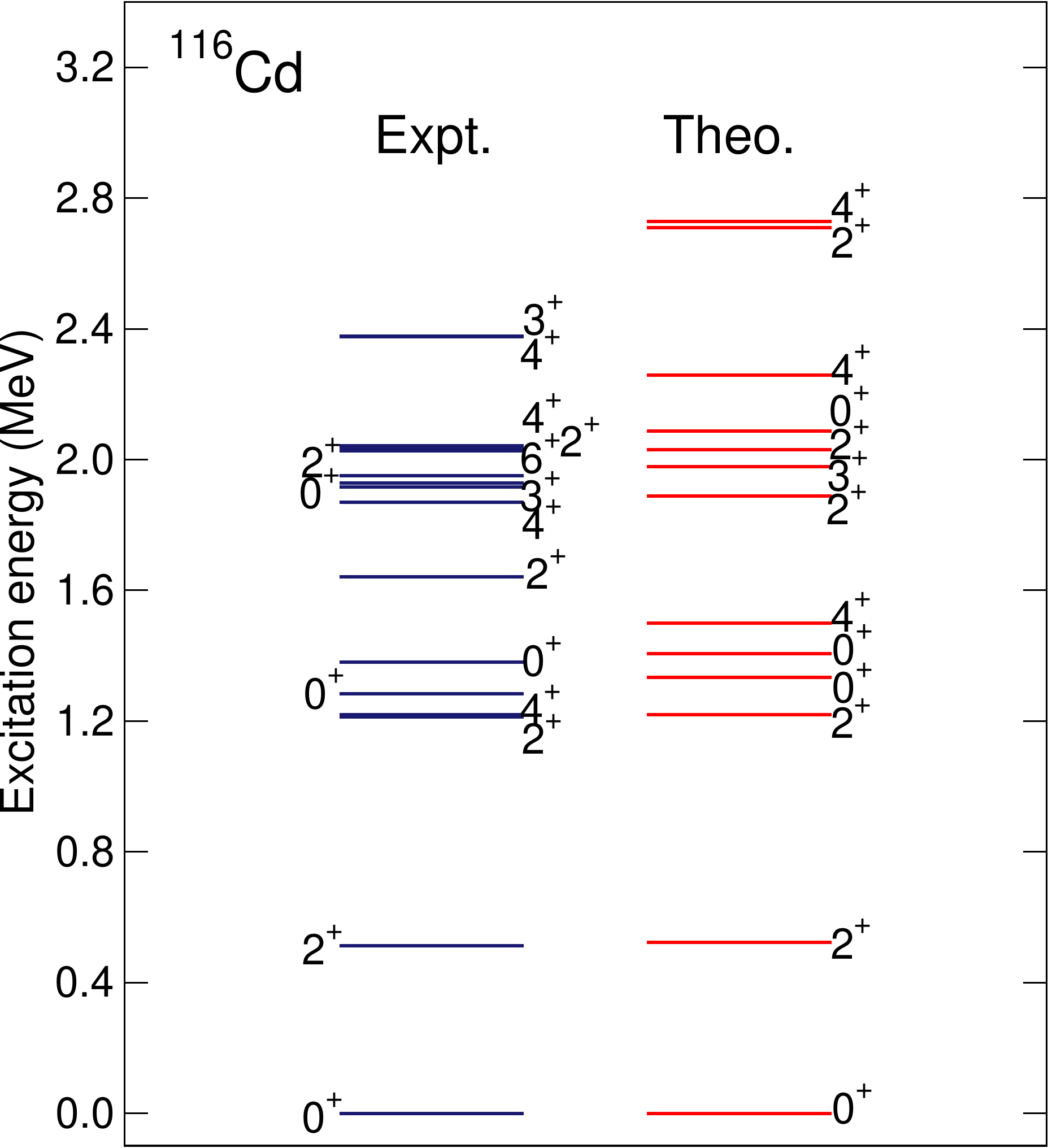} \\
\end{tabular}
\caption{(Color online) Experimental and predicted excitation spectra
 for $^{108-116}$Cd.} 
\label{fig:spectra}
\end{center}
\end{figure*}

\begin{figure}[htb!]
\begin{center}
\includegraphics[width=0.6\linewidth]{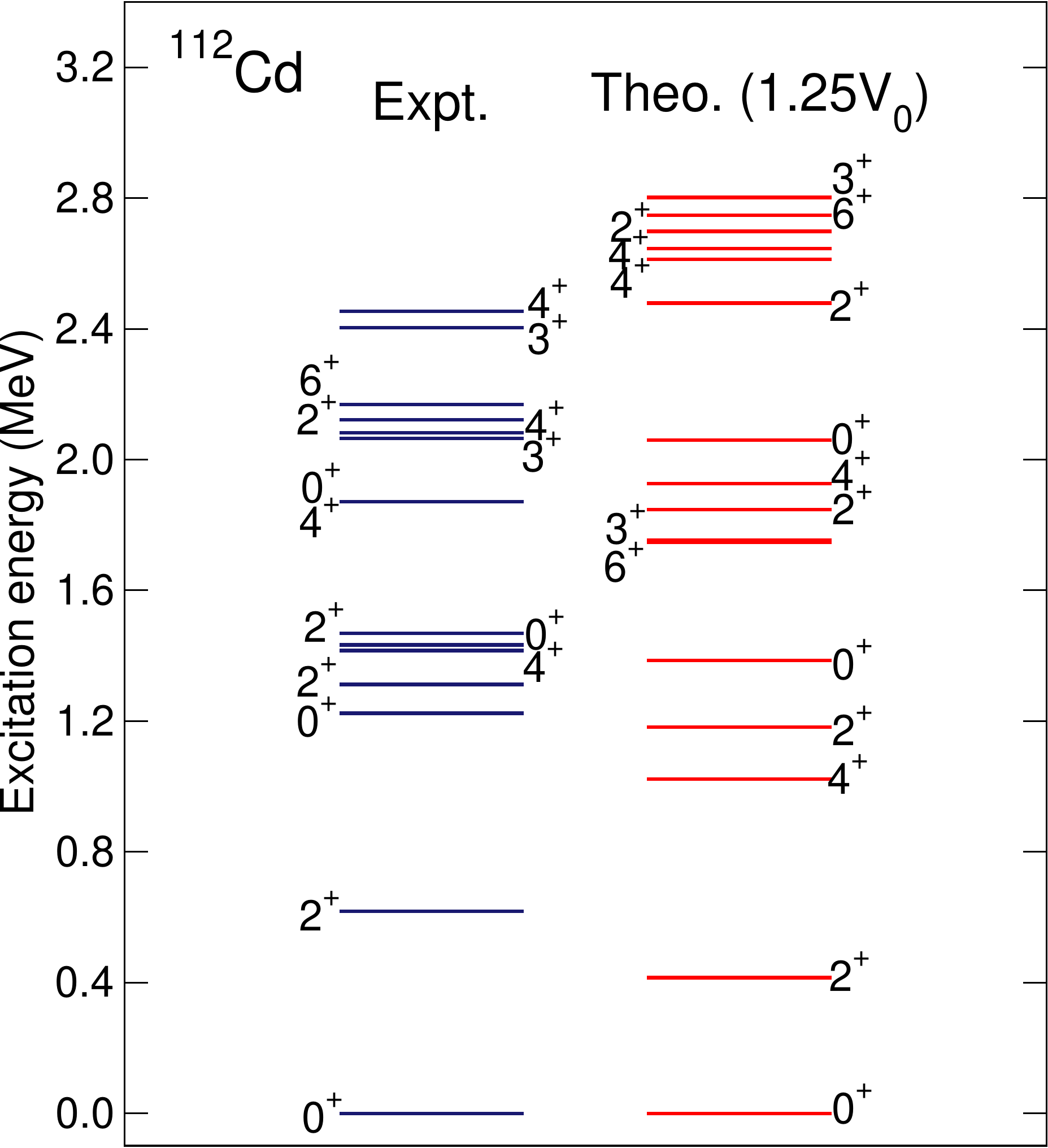} 
\caption{(Color online) Same as Fig.~\ref{fig:spectra}, but for the
 calculation on $^{112}$Cd based on the pairing strength increased by 25
 \%} 
\label{fig:spectra-pairing}
\end{center}
\end{figure}

Figure~\ref{fig:spectra} shows the comparison of the predicted and
experimental excitation spectra for $^{108-116}$Cd. The experimental
data are taken from \citep{BLA00}, \citep{GUR12}, \citep{LAL15}, \citep{BLA12},
\citep{BLA10} for $^{108-116}$Cd respectively. Shown are the
experimental levels up to an energy of about 2.5 MeV and maximally five
states are shown for a given spin $J^{+}$.  The corresponding
theoretical levels are given up to an energy of about 2.8 MeV. The
predicted energies of the intruder states can be seen in those of the
third or higher 0$^{+}$ states (see, Table~\ref{tab:frac}). In
Fig.~\ref{fig:spectra}, the semi-microscopic calculations 
based on the energy surface  do predict the approximately right
excitation energies and the dependence on neutron number with the lowest
$0^+_3$ energy at mid-shell in $^{114}$Cd. 
The description of the spacing between both 0$^{+}$ states is
overestimated in $^{108,110}$Cd, but agrees quite well with experiment
in $^{112-116}$Cd. 
However, phenomenological calculations assign
in $^{114}$Cd the experimental $0^{+}_2$ state to the intruder
configuration \citep{027}. Therefore, the energy difference between
intruder and normal states is generally overestimated. 
The predicted normal states are systematically too 
deformed as can be seen by the energies of the first 2$^{+}$, 4$^{+}$
and especially 6$^{+}$ states. Experimentally states are observed with
more vibrational energies although there are serious problems with the
electric quadrupole transitions \citep{024}. 
The description of the
energies of the 2$^{+}$ intruder state is not good. Phenomenological
calculations, i.e. \citep{152Jolie22_not_listed, 143Jolie09}, identified
in $^{110-114}$Cd the 2$^{+}_3$ state as the intruder state, while the
present calculation yield as main component the much higher lying
2$^{+}_4$ state. In contrast to the normal states the spacing between
the 0$^{+}$ and 2$^{+}$ intruder states is too large.
As expected the
semi-microscopic predictions have problems to describe the increasing
level densities above 2 MeV.  

\begin{table}
 \begin{center}
\caption{\label{tab:frac} Fraction (in units of percent) of the
  intruder configuration in the lowest five $0^+$ and three $2^+$ wave
  functions of the considered Cd nuclei.}
  \begin{tabular}{cccccc}
\hline\hline
 & $^{108}$Cd & $^{110}$Cd & $^{112}$Cd & $^{114}$Cd & $^{116}$Cd \\
\hline
$0^+_1$ & 1 & 2 & 3 & 5 & 4 \\
$0^+_2$ & 2 & 3 & 6 & 29 & 3\\
$0^+_3$ & 5 & 26 & 55 & 52 & 82 \\
$0^+_4$ & 5 & 74 & 46 & 26 & 20 \\
$0^+_5$ & 91 & 3 & 7 & 8 & 40 \\
$2^+_1$ & 1 & 2 & 3 & 4 & 4 \\
$2^+_2$ & 3 & 5 & 8 & 16 & 17 \\
$2^+_3$ & 1 & 2 & 4 & 9 & 74 \\
$2^+_4$ & 6 & 55 & 81 & 74 & 10 \\
\hline\hline
  \end{tabular}
 \end{center}
\end{table}

In order to investigate the influence of the density-dependent pairing
interaction we have performed the IBM-2 calculations using the energy surface shown
in Fig.~\ref{fig:pes-pairing}. The corresponding excitation spectrum is shown in
Fig.~\ref{fig:spectra-pairing}. As there is no second minimum we only have
states corresponding to 
the normal states. This is clearly not the case experimentally as
reflected i.e. by the excited 0$^{+}$ states. On the other hand the
comparison for the energies of the normal states is substantially
improved i.e. for the 6$^{+}$ states. 

The theoretical B(E2;J$^+_i \rightarrow$ J$^+_f$) values were calculated
for $^{108-116}$Cd using the transition operator given in Eq.~(\ref{eq:e2}) with
fixed values $e^\nu_1=e^\pi_1=0.084\,eb$ and $e^\nu_3=e^\pi_3=0.113\,eb$, which are taken from
\citep{152Jolie22_not_listed}. Table~\ref{tab:Be2} compares them in
Weisskopf units (W.u.) with the experimental data, when available. In view of the absence
of fitting the data, there is good agreement and changing trends are
well described. However, there are exceptions, mostly involving the
0$^{+}_{2}$, 0$^{+}_{3}$, 2$^{+}_{2}$ and 2$^{+}_{3}$ states as could be
expected from the different nature of these states compared to the
phenomenological IBM-2 calculations.  As an example the B(E2;0$^+_2
\rightarrow$ 2$^+_1$) are ten times underpredicted in $^{112,114}$Cd and
ten times overpredicted in $^{116}$Cd. This confirms that these states,
which are the lowest where normal and intruder states mix, are poorly described. 
Another reason could be, again, that at the SCMF level the prolate
minimum, from which the normal states are mainly constructed, is predicted to be
too deformed (see, Fig.~\ref{fig:pes}). 

When using the increased pairing for
$^{112}$Cd and assuming that the 0$^{+}_{3}$ and 2$^{+}_{3}$ are outside
the model space the B(E2;0$^+_2 \rightarrow$ 2$^+_1$) = 15 W.u. is still
underpredicted by a factor three. 
We note that there is, overall, no striking difference between the
predictions of the $B$(E2) rates with different pairing strengths. 
Exceptions are perhaps $B(E2; 2^+_4 \rightarrow 0^+_3)=2.3$ and 
$B(E2; 2^+_5 \rightarrow 0^+_3)=15$ W.u., which were obtained with the
increased pairing and are by about a factor ten different from those results with
the pairing strength of $V_0=1000$ MeV fm$^3$ (see,
Table~\ref{tab:Be2}). 
Not surprisingly, such a difference occurs due to whether or not the intruder
configuration is included. For instance, the transition $2^+_4
\rightarrow 0^+_3$ is between the states mainly made of the intruder
components in the configuration mixing calculation (see,
Table~\ref{tab:frac}), while it is between normal states with the
pairing strength of $V_0=1250$ MeV fm$^3$.

\begin{table*}
 \begin{center}
\caption{\label{tab:Be2}Comparison between experimental and theoretical B(E2;J$^+_i
  \rightarrow$ J$^+_f$) values in Weisskopf units.}
  \begin{tabular}{cccccccccccc}
\hline\hline
\multirow{2}{*}{$J_i^+$} & \multirow{2}{*}{$J_f^+$} & \multicolumn{2}{c}{$^{108}$Cd} &  \multicolumn{2}{c}{$^{110}$Cd} &
   \multicolumn{2}{c}{$^{112}$Cd} &  \multicolumn{2}{c}{$^{114}$Cd} &
   \multicolumn{2}{c}{$^{116}$Cd} \\
\cline{3-4}
\cline{5-6}
\cline{7-8}
\cline{9-10}
\cline{11-12}
& & Exp\footnote{all from \citep{BLA00}}&Theory &Exp\footnote{from P.E. Garrett  et al.,
  \citep{019} except italic  from \citep{GUR12}}&Theory&Exp\footnote{from \citep{LAL15}, except italic  from P.E. Garrett et al. \citep{021}}&Theory&Exp\footnote{all from \citep{BLA12}, except italic  from M. D\'{e}l\`{e}ze et al. \citep{152Jolie22_not_listed}}&Theory &Exp\footnote{all from \citep{BLA10} except italic  from M. Kadi  et al.\citep{031}}&Theory\\
\hline
2$_{1}$	&0$_{1}$&26.6(3)&29&${\it 27.0(8)}$&33	&30.3(2)&39	&31.1(19)&46&33.5(12)&36\\
0$_{2}$&2$_{1}$&-&1.4&$<$40&2.8&51(14)&4.5&27.4(17)&2.9&0.79(22)&9.5\\
2$_{2}$&0$_{1}$&1.8(3)&1.1&0.68(14)&1.7	&0.65(11)&2.4&0.48(6)&3.2&1.11(18)&	1.9\\
2$_{2}$	&2$_{1}$&17(5)&	6&19(4)or ${\it 30(5)}$ &	11&39(7)&18	&22(6)&21&25(10)&27\\
2$_{2}$	&0$_{2}$&-&1.7&	${\it 1.35(20)}$&1.2&-&	2.6&3.4(7)	&11&-&1.7\\
4$_{1}$&2$_{1}$&41(6)&39&${\it 42(9) }$	&47&63(8)&55	&62(4)&	65&56(14)&51\\
0$_{3}$&2$_{1}$	&-&0.003&$<$7.9&0.10&0.0121(17)&0.82&0.0026(4)&4.4	&30(6)&	1.6\\
0$_{3}$	&2$_{2}$&-&13&$<$1680&29&99(16)&42&127(16)&39&-&96\\
2$_{3}$	&0$_{1}$&-&0.02	&0.28(4)&0.051&	0.88(17)&0.085&0.33(4)&0.072&1.11(18)&0.25\\
2$_{3}$	&2$_{1}$&-&0.02	&${\it 0.7^{+3}_{-4}}$	&0.068	&${\it 0.12(7)}$&0.14&$<$0.045&0.17&6.2$^{+22}_{-26}$&0.0083\\
2$_{3}$&0$_{2}$	&-&16&29(5)&20&	120(50)	&25	&65(9)	&32&-&	2.8\\
2$_{3}$&2$_{2}$&-&0.17&	$<$8&0.46	&-&0.79&-&	0.22&	-& 7.8\\
2$_{3}$&0$_{3}$	&-&0.56&-&0.43&	-&0.98&-&	1.9&${\it 86^{+24}_{-30}}$&76\\
3$_{1}$&2$_{1}$&-&1.5&	0.85(25)&2.5&1.8(5)&3.3&-&	4.2&${\it 2.6(7) }$&2.0\\
3$_{1}$&2$_{2}$&-&30&22.7(69)&38&64(18)	&47	&-&55	&${\it 61(17) }$& 39\\
3$_{1}$&4$_{1}$&-&3.9&	2.4$^{+9}_{-8}$	&6.8&25(8)&10	&-&12&${\it 18(10)}$&11\\
3$_{1}$&2$_{3}$&-&2.3&	$<$5&1.9&	-&1.6&-&1.9&-&3.6\\

4$_{2}$&2$_{1}$&-&0.035&	0.14(6)	&0.083	& -	&0.14&0.50(5)&0.32&${\it 3.0(7)}$&0.22\\
4$_{2}$&2$_{2}$&-&15&	22(10)&	23& - &31	&32(4)	&45&${\it 230(130)}$&44\\
4$_{2}$&4$_{1}$&-&4.8&	10.7$^{+49}_{-48}$&8.6&	- &13&17(6)&16&${\it 150(90) }$&18\\
4$_{2}$&2$_{3}$	&-&1.4	&$<$0.5&0.98& - &0.79&119(12)&5.9	&-&31\\

2$_{4}$&0$_{1}$	&-&0.011&-&0.10&	0.017(5)&0.19&0.19(4)&0.28&${\it 0.13(4) }$&0.011\\
2$_{4}$&2$_{1}$&-&0.021&${\it 0.28^{+6}_{-10}}$&0.032&	2.2(6)&	0.0011&	0.84(17)&0.059&${\it 1.23(44) }$&0.16\\
2$_{4}$&0$_{2}$&-&0.28&	-&0.88&	5.3(15)&2.2&42(9)&	15&-& 21\\
2$_{4}$&2$_{2}$&-&0.0011&-&1.6&	${\it <2.8}$&2.7&5.6(11)&5.8&${\it 31(9) }$&0.27\\
2$_{4}$&0$_{3}$&-&13&-&	35&25(7)&62&34(8)&	68&-&	5.6\\
2$_{4}$&2$_{3}$&-&0.87&-&0.97&	-&0.83&70(40)&16&-&	21\\
0$_{4}$&2$_{1}$&-&0.0049&-&0.32&${\it <1.4}$&0.50&2.4(6)&0.44	&-&0.25\\
0$_{4}$&2$_{3}$&-&1.3&	-&0.17&	${\it <23}$&1.0&18(6)&10&	-&64\\

4$_{3}$&2$_{1}$&-&0.0044&0.14(4)&0.0019&0.9(3)&0.012&-&0.057&-&0.11\\
4$_{3}$&2$_{2}$&-&0.0000&1.2(4)	&1.7& 58(17) &	2.8&-&4.8&	-&0.64\\
4$_{3}$&4$_{1}$&-&0.0038&1.8$^{+10}_{-15}$&0.032& 24(8) &0.13&-	&0.35&	-&0.67\\
4$_{3}$&2$_{3}$&-&18&115(35)&0.99& 59(20) &0.14&-	&3.2&-	&86\\

6$_{1}$&4$_{1}$	&-&39&62(18)&	49&-&59&119(15)&	72&${\it 110(46)}$&58\\
6$_{1}$&4$_{3}$&-&0.47&	36(11)&	0.11&-&	0.064&-&0.011&-&0.90\\
2$_{5}$&0$_{1}$&-&0.0036&-&0.093&${\it 0.136(16) }$&0.038&0.08(3)&0.012& -&0.0048\\
2$_{5}$&2$_{1}$	&-&0.021&3.2(3)	&0.077	&${\it 0.060^{+27}_{40}}$&0.063&-&0.19&	-&0.079\\
2$_{5}$&2$_{5}$	&-&0.055&0.7$^{+5}_{-6}$&0.087	&-&0.0098&	$<$1.9&0.018&-& 0.014\\
2$_{5}$&0$_{3}$&-&0.99&	24.2(22)&0.15&	-&0.42&17(5)	&0.15	&-&0.71\\
2$_{5}$&2$_{3}$&-&4.2&	$<$5&0.23&${\it 22^{+6}_{-19}}$&	0.37&-&0.71&-&13\\
8$_{1}$&6$_{1}$&-&34&80(22)\footnote{assuming that the second $8^+$ state corresponds to the first theoretical $8^+$ state.}&45&-&	58&	${\it 86(28) }$	&73&-&	61\\
6$_{2}$&4$_{2}$	&-&23&	-&33&$<$77&49&${\it 129(42) }$	&68&-& 88\\
\hline
\end{tabular}
 \end{center}
\end{table*}

By using the E0 transition operator in Eq.~(\ref{eq:e0}) we have
calculated the theoretical $\rho^2({\rm E0}; J_i\rightarrow J_f)$ values: 
\begin{eqnarray}
 \label{eq:rhoe0}
\rho^2({\rm E0}; J_i\rightarrow J_f) = \frac{Z^2}{e^2R^4}|\langle J_f|\hat T^{E0}|J_i\rangle|^2,
\end{eqnarray}
where $R=1.2A^{1/3}$ fm, and for the parameters in the E0 operator
we used the fixed values $\beta_{\nu,1}=\beta_{\nu,3}=0.10$ fm$^2$,
$\beta_{\pi,1}=\beta_{\pi,3}=0.60$ fm$^2$, taken from
Ref.~\citep{Giannatiempo1991}. We have also set 
$\gamma_{\nu,1}=\gamma_{\nu,3}=0$ fm$^2$. 
In Table~\ref{tab:e0} the theoretical $\rho^2$(E0) values are compared
with the available experimental data. 
The large experimental error bars make a clear comparison in $^{110}$Cd
difficult. However, in $^{112}$Cd were the values are better defined the
agreement is very good with one exception (the $0^+_3$  to $0^+_1$ 
transition). In $^{114}$Cd, where the mixing is the strongest the E0
transitions are instead very poorly described. 

\begin{table}[htb]
\caption{\label{tab:e0} Comparison between experimental and theoretical
 $\rho^2({\rm E0}; J^+_i\rightarrow J^+_f)$ values. The
 experimental $\rho^2$(E0) values are not known for $^{108}$Cd and $^{116}$Cd.}
 \begin{center}
  \begin{tabular}{ccccc}
\hline\hline
\multirow{2}{*}{} & \multirow{2}{*}{$J_i^+$} & \multirow{2}{*}{$J_f^+$} &
   \multicolumn{2}{c}{$\rho^2$(E0)$\times 10^3$} \\
\cline{4-5}
 & & & Exp & Theory \\
\hline
$^{110}$Cd & $0^{}_2$ & $0^{}_1$ & $<$31(5)\footnotemark[1] & 37 \\
           & $0^{}_3$ & $0^{}_1$ & $<$11\footnotemark[2] & 1.1 \\
           & $2^{}_2$ & $2^{}_1$ & 20(15)\footnotemark[3] & 1.1 \\
           & $2^{}_3$ & $2^{}_1$ & 9(8)\footnotemark[1] & 26 \\
           & $4^{}_3$ & $4^{}_1$ & 106$^{+98}_{-91}$\footnotemark[1] & 0.44 \\
$^{112}$Cd & $0^{}_2$ & $0^{}_1$ & 34(9)\footnotemark[4] & 36 \\
           & $0^{}_3$ & $0^{}_1$ & 0.87(5)\footnotemark[4] & 8.6 \\
           & $0^{}_3$ & $0^{}_2$ & 10.7(6)\footnotemark[4] & 12 \\
           & $2^{}_3$ & $2^{}_1$ & 31(20)\footnotemark[3] & 27 \\
$^{114}$Cd & $0^{}_2$ & $0^{}_1$ & 19(2)\footnotemark[4] & 12 \\
           & $0^{}_3$ & $0^{}_1$ & 1.83(13)\footnotemark[4] & 44 \\
           & $0^{}_3$ & $0^{}_2$ & 0.65(5)\footnotemark[4] & 100 \\
           & $0^{}_4$ & $0^{}_1$ & 0.9(4)\footnotemark[4] & 8.8 \\
           & $2^{}_2$ & $2^{}_1$ & $<$28\footnotemark[3] & 0.25 \\
           & $2^{}_3$ & $2^{}_1$ & 38(5)\footnotemark[5] & 22 \\
           & $2^{}_3$ & $2^{}_2$ & 22(6)\footnotemark[5] & 1.1 \\
           & $2^{}_4$ & $2^{}_2$ & $<$20\footnotemark[5] & 57 \\
           & $3^{}_2$ & $3^{}_1$ & $<$130\footnotemark[5] & 35 \\
           & $4^{}_2$ & $4^{}_1$ & 67(10)\footnotemark[5] & 0.38 \\
\hline\hline
  \end{tabular}
\footnotetext[1]{Reference \citep{Jigmeddorj2016}}
\footnotetext[2]{Reference \citep{garrett2018privcomm}}
\footnotetext[3]{Reference \citep{Giannatiempo1991}}
\footnotetext[4]{Reference \citep{kibedi2005}}
\footnotetext[5]{Reference \citep{garrett2016}}
 \end{center}
\end{table}

\section{Conclusion}

Based on constrained self-consistent mean-field calculations,
deformation energy surfaces were calculated for the even-even
$^{108-116}$Cd isotopes. 
The energy surfaces yielded both a prolate and a minor oblate minimum
which were consider to be associated to proton 0p-0h normal excitations and
2p-2h intruder excitations. 
They were used to fit the parameters of an
IBM-2 Hamiltonian, which involves normal and intruder states and their
mixing. The resulting energy spectra, and B(E2) and $\rho^2$(E0) values were compared with
experiment. An overall reasonable agreement was found in view of the
fact that no additional fitting to the experiment was done. 
The calculations describe the energy dependence of the lowest intruder state with
the expected minimum at mid shell, but predict these states at higher
energies than phenomenological calculations. Therefore the mixing of the
lowest two intruder states with the normal states is not reproduced
correctly leading to large discrepancies for the B(E2) and $\rho^2$(E0) values. Moreover,
the normal states are predicted to be too deformed. We
also studied the effect of the pairing interaction  in the case of
$^{112}$Cd and found that an increase of the strength leads to a
disappearance of the intruder states. 

\begin{acknowledgments}
The authors thank N. Gavrielov for useful comments. 
KN is grateful to the University of Cologne for their kind hospitality
 and financial support. 
He acknowledges support by the QuantiXLie Centre of Excellence, a project
co-financed by the Croatian Government and European Union through the
European Regional Development Fund - the Competitiveness and Cohesion
Operational Programme (Grant KK.01.1.1.01.0004).
JJ acknowledges financial support from the Interuniversity Attraction Poles Program
of the Belgian State-Federal Office for Scientific and Cultural Affairs
(IAP Grant P7/12) and from GANIL (Caen France) were part of this work was done. 
\end{acknowledgments}


\end{document}